\documentclass[letterpaper,twocolumn,amsmath,amssymb,pre,aps,10pt]{revtex4-1}
\usepackage{graphicx} 
\usepackage{color}
\usepackage{nicefrac} 

\usepackage{xargs}                      
\usepackage[pdftex,dvipsnames]{xcolor}  
\usepackage[colorinlistoftodos,prependcaption,textsize=normalsize]{todonotes}
\usepackage{mdframed}

\definecolor{dark-gray}{gray}{0.10}
\definecolor{light-gray}{gray}{0.70}

\newcommandx{\jpcom}[2][1=inline]{\todo[linecolor=gray,backgroundcolor=light-gray,bordercolor=dark-gray,#1]{\textbf{Jordan says:} #2} }
\begin{document}

\title{Stochastic Approximation Monte Carlo with a Dynamic Update
Factor
}

\author{Jordan K. Pommerenck} \author{Tanner T. Simpson}
\author{Michael A. Perlin} \author{David Roundy}
\affiliation{Department of Physics, Oregon State University,
  Corvallis, OR 97331}

\begin{abstract}
  We present a new Monte Carlo algorithm based on the Stochastic
  Approximation Monte Carlo (SAMC) algorithm for directly calculating
  the density of states. The proposed method is Stochastic
  Approximation with a Dynamic update factor (SAD)
  which dynamically adjusts the update factor $\gamma_t$ during the course of
  the simulation. We test this method on the square-well fluid and
  the 31-atom Lennard-Jones cluster and
  compare the convergence behavior of several
  related
  Monte Carlo methods. We find that both the SAD and $1/t$-Wang-Landau ($1/t$-WL)
  methods rapidly converge to the
  correct density of states without the need for the user to specify an
  arbitrary tunable parameter $t_0$ as in the case of SAMC.  SAD requires
  as input the temperature range of interest, in contrast to
  $1/t$-WL, which requires that the user identify the interesting range
  of energies.
  The convergence of the $1/t$-WL method is very sensitive to the energy
  range chosen for the low-temperature heat capacity of the
  Lennard-Jones cluster.
  Thus, SAD is more powerful in the common case in which the range
  of energies is not known in advance.
\end{abstract}

\maketitle

\section{Introduction}
Over the past several decades, a number of flat histogram Monte Carlo simulation
algorithms have been developed which calculate the thermodynamic properties of
various systems over a range of temperatures.  This development began with the
original histogram method, which used a single canonical Monte Carlo simulation
to predict properties for nearby temperatures~\cite{ferrenberg1988new}.  For
large systems, this approach is limited to a narrow temperature range because a
single canonical simulation explores only a small range of energies. Berg and
Neuhaus developed multicanonical methods that introduced a weight function to
enable flat histogram sampling which improved the exploration of configuration
space and allowed the system to overcome free energy
barriers~\cite{berg1991multicanonical, berg1992multicanonical}. These works led
to an explosion in the interest of ``flat'' (or ``broad'') histogram
methods~\cite{penna1996broad, penna1998broad, swendsen1999transition,
wang2001determining, wang2001efficient, landau2004new, schulz2003avoiding,
yan2003fast, trebst2004optimizing, belardinelli2007wang, belardinelli2007fast,
belardinelli2008analysis, belardinelli2014intrinsic, singh2012density,
zhou2008optimal, schneider2017convergence, liang2006theory, liang2007stochastic,
liang2009improving, werlich2015stochastic, kim2006statistical,
kim2007statistical, kim2009replica, junghans2014molecular}, which explore a
wider range of energies. In addition to simulating a range of temperatures, in
contrast with low-temperature canonical Monte Carlo, these approaches
avoid being trapped in a local energy minimum.

Wang and Landau introduced the most widely used flat histogram algorithm (WL)
that uses an update factor and a statistical histogram to compute the density of
states of a given system~\cite{wang2001determining, wang2001efficient}.  While
the method is incredibly powerful, it has a few disadvantages. The most
significant disadvantage is that the method requires the user to select the
range of energies to be studied~\cite{landau2004new, wang2001efficient,
schulz2003avoiding, yan2003fast}. This requirement adds an additional
hurdle to its application to systems for which the interesting range of energies
is not known \emph{a priori}.  The simulation violates detailed balance, albeit
briefly as the size of the violation decreases with time, which complicates
convergence analysis.  In fact, the error in a WL computation has been
demonstrated to saturate at a non-zero value~\cite{yan2003fast}, i.e. the method
does not converge to the true density of states~\cite{belardinelli2007wang,
belardinelli2007fast, belardinelli2008analysis, belardinelli2014intrinsic,
singh2012density, zhou2008optimal}.

Belardinelli and Pereyra demonstrated that allowing the update factor to
decrease faster than $1/t$ leads to nonconvergence~\cite{belardinelli2007wang},
where $t$ corresponds to the number of moves.
This leads to their $1/t$-WL algorithm which ensures that the error continues to
decrease asymptotically as $1/\sqrt{t}$, which they demonstrated avoids
error saturation and asymptotically approaches the true density of
states~\cite{belardinelli2008analysis}. Zhou \emph{et al.} further confirmed that the
original WL algorithm never converges exponentially and successfully bounded the
statistical error between $t^{-\frac12}$ and $1/t$~\cite{zhou2008optimal}.
Schneider \emph{et al.} outline minor refinements to the $1/t$-WL algorithm
including scaling the update factor with the number of energy
bins~\cite{schneider2017convergence}.

Liang began to consider Stochastic Approximation as a mathematical
generalization of the WL approach, with convergence that could be mathematically
proven~\cite{liang2006theory}. In 2007, Liang \emph{et al.}
developed Stochastic Approximation Monte Carlo (SAMC)~\cite{liang2007stochastic,
liang2009improving}, and proved its convergence, although the method still has a
system specific user-defined parameter which must be tuned when applying this
algorithm to a new system.  In contrast to the WL-based methods, the SAMC method
does not require users to determine the energy range of interest \emph{a
priori}. Werlich \emph{et al.} proposed the introduction of an additional tuning
parameter into SAMC~\cite{werlich2015stochastic}.

Kim \emph{et al.} introduced Statistical Temperature Monte Carlo (STMC) and the
related Statistical Temperature Molecular Dynamics (STMD) which is an
adaption of the WL method that approximates the entropy (or natural logarithm of the density of
states) as a piecewise linear function, which improves convergence for systems
with a continuously varying energy~\cite{kim2006statistical,
kim2007statistical}. STMC applied to WL requires a temperature range be
specified rather than an energy range.  Kim \emph{et al.} extended this work as
Replica Exchange Statistical Temperature Monte Carlo (RESTMC), which uses
replica exchange of multiple overlapping STMC simulations to improve
convergence~\cite{kim2009replica}. More recently, Junghans \emph{et al.} have
demonstrated a close connection between metadynamics, which was introduced by
Laio and Parinello~\cite{laio2002escaping}, and WL-based Monte Carlo methods,
with STMD forging the connection~\cite{junghans2014molecular}.

In this work, we have developed an improved algorithm based on SAMC that does
not require an array of non-physical, user-defined inputs and therefore should
be easily applicable to any system. The method (like STMC above) does require
the user to define a temperature range of interest ($T_\text{min}$ to
$T_\infty$) which we explain in Section~\ref{sec:sad}.  We call this method SAD (Stochastic Approximation with a Dynamic
update factor), and will discuss it in detail in the methods section. We compare
its convergence properties with three existing flat histogram methods: WL,
$1/t$-WL, and SAMC.

In this work, we compare four flat histogram methods.  We outline the
general workings of each algorithm that we developed in detail while
summarizing algorithms that were developed in other works.  The
following methods are discussed and simulated for the square-well
fluid and the 31-atom Lennard-Jones cluster: Wang-Landau (WL),
$1/t$-Wang-Landau ($1/t$-WL), Stochastic Approximation Monte Carlo
(SAMC), and SAD.

\section{Flat histogram methods}\label{sec:histogram}
The goal of flat histogram methods (also called \emph{broad histogram}
or \emph{multicanonical} methods) is to simulate each energy with
similar accuracy so as to accurately determine the density of states
over a broad range of energies---and thus to determine the thermodynamic
quantities such as heat capacity or internal energy
over a broad range of temperatures.
Properties that require more information---such as a spatial
correlation function or a response function---can still be computed
for any temperature, provided statistics are collected for each
individual energy, which can then be reweighted for any
temperature~\cite{panagiotopoulos1998phase, panagiotopoulos2000monte,
errington2003direct}.

All the flat histogram Monte Carlo methods begin with randomly chosen
``moves'' which change the state of the system and must satisfy
detailed balance.  Each algorithm differs in how it determines the
probability of accepting a move and in what additional statistics must
be collected in order to decide on that probability.

Flat histogram methods calculate the density of states $D(E)$ for a discrete set
of energies~\cite{wang2001determining, dayal2004performance, troyer2003flat,
trebst2004optimizing}. Therefore, energy binning becomes an important
consideration for systems with a continuum of possible energies.  Energy bins
are typically of uniform size for the entire energy
continuum~\cite{fasnacht2004adaptive}. Some methods such as
AdaWL~\cite{koh2013dynamically} employ a tunable mechanism for controlling the
binning for low entropic states in order to ensure the exploration of all
energies.  The method introduced in this paper is designed to scale appropriately
as bin size is changed.

In this section we will introduce four closely related flat histogram
methods each of which rely on a weight function $w(E)$.  In these
algorithms, the probability of accepting a move is given by
\begin{equation}
	\mathcal{P}(E_\text{old} \rightarrow E_\text{new})
	= \min\left[1,\frac{w(E_\text{old})}{w(E_\text{new})}\right]
\end{equation}
which biases the simulation in favor of energies with low weights.  A set of
weights that are proportional to the density of states $D(E)$ of the system will
result in an entirely flat histogram, which means that the weights should
converge to being proportional to the density of states.  The natural logarithm
of the weight is typically stored, since the density of states will often vary
over a few hundred orders of magnitude. In the microcanonical ensemble, the
entropy is defined as $S(E) \equiv \ln(D(E))$ (where $k_B \equiv 1$), the
logarithm of the weight is an approximation of the entropy.

Each approach uses a random walk in energy space to estimate the density of
states.  The core of these approaches is to update the weights at each step of
the simulation
\begin{equation}
	\ln{w_{t+1}(E)}=\ln{w_{t}(E)}
	+\gamma_t
\end{equation}
where $t$ is the number of the current move, $\gamma_t$ is an update factor
that varies over the simulation, and $E$ is the current energy.  This update
causes the random walk to avoid energies that have been frequently sampled,
enabling a rapid exploration of energy space. This approach violates detailed
balance, due to the acceptance probabilities changing with each move, but the
severity of this violation decreases as we decrease $\gamma_t$.  The four
methods differ primarily in how they schedule the decrease of $\gamma_t$.

\subsection{Wang-Landau}

The Wang-Landau approach~\cite{wang2001efficient,wang2001determining,
landau2014guide} begins with $\gamma^{\text{WL}}_{t=0}=1$, and then decreases
$\gamma^{\text{WL}}$ in discrete stages.  When the energy histogram is
sufficiently flat, $\gamma^{\text{WL}}$ is decreased by a specified factor of
$\frac12$.  The flatness is defined by the ratio between the minimum value of
the histogram and its average value.  When this flatness reaches a specified
threshold (typically 0.8), the $\gamma^{\text{WL}}$ value is decreased and the
histogram is reset to zero.  This approach requires that the energy range of
interest be known in advance, and difficulties can occur with this flatness
criteria due to the fact that some energies in this energy range might never be
sampled~\cite{troster2005wang}.  The entire process is repeated until
$\gamma^{\text{WL}}$ reaches a desired cutoff.

The Wang-Landau approach thus has three parameters that need be
specified: the factor by which to decrease $\gamma^{\text{WL}}$ when flatness is
achieved, the flatness criterion, and the cutoff that determines when
the computation is complete.  In addition, an energy range (or in
general, a \emph{set} of energies) must be supplied, so that the
flatness criterion can be defined.

\subsection{$1/t$-Wang-Landau}

The $1/t$-WL algorithm ensures convergence by preventing the $\gamma_t$ factor
from dropping below $N_S/t$~\cite{belardinelli2008analysis,
schneider2017convergence}. The method follows the standard WL algorithm with two
modifications.  Firstly, the histogram is considered flat, and $\gamma_t$ is
decreased by a factor of two, when every energy state has been visited once,
i.e. when the WL ``flatness'' becomes nonzero. Secondly, when
$\gamma^{\text{WL}} < N_S/t$ at time $t_0$, the algorithm switches to use
$\gamma_t = N_S/t$ for the remainder of the simulation:
\begin{align}
  \gamma_t^{1/t\text{-WL}} = \begin{cases}
     \gamma^{\text{WL}}_t & \gamma^{\text{WL}}_t > \frac{N_S}{t} \\
     \frac{N_S}{t} & t \ge t_0
 \end{cases}
\end{align}
where $t$ is the number of moves, $\gamma^{\text{WL}}_t$ is the Wang-Landau update factor
at move $t$, and $N_S$ is the number of energy bins.

\subsection{SAMC}
The Stochastic Approximation Monte Carlo (SAMC) algorithm addresses the lack of
convergence of Wang-Landau's approach with a simple schedule by which the update
factor $\gamma_t$ is continuously decreased~\cite{liang2007stochastic,
werlich2015stochastic, schneider2017convergence}.  The update factor is defined
in the original implementation~\cite{liang2007stochastic} in terms of an tunable
parameter $t_0$,
\begin{align}
\gamma_{t}^{\text{SA}} =\frac{t_0}{\max(t_0,t)}\label{eq:1}
\end{align}
where as above $t$ is the number of moves that have been attempted.
SAMC offers extreme simplicity, combined with is proven convergence.
Provided the update factor satisfies
\begin{align}
\sum_{t=1}^\infty \gamma_{t} = \infty \quad\textrm{and}\quad
\sum_{t=1}^\infty \gamma_{t}^\zeta < \infty
\end{align}
where $\zeta > 1$, Liang has proven that the weights converge to the true
density of states~\cite{liang2006theory, liang2007stochastic,
liang2009improving}.  In addition, the energy range need not be known \emph{a priori}.
The time to converge depends only on the choice of parameter $t_0$.
Unfortunately, there is no prescription for finding an acceptable value for
$t_0$, and while the algorithm formally converges, for a poor choice of $t_0$
that convergence can be far too slow to be practical. Liang \emph{et al.} give a
rule of thumb in which $t_0$ is chosen in the range from $2N_S$ to $100N_S$
where $N_S$ is the number of energy bins~\cite{liang2007stochastic}. Schneider
\emph{et al.} found that for the Ising model this heuristic is helpful for small
spin systems, but that larger systems require an even higher $t_0$
value~\cite{schneider2017convergence}.  We will describe below one case we
examined, in which $t_0$ needs to be as much as two orders of magnitude higher
than the rule of thumb of $100N_S$ in order to converge in $10^{12}$ moves.

Werlich \emph{et al.} proposed scaling the SAMC $\gamma_t^{\text{SA}}$
by a factor $\gamma_0$~\cite{werlich2015stochastic}.  While this may result in
an improved rate of convergence, it adds yet another parameter that must be
empirically determined, and we have not explored this additional degree of
freedom.

\subsection{SAMC convergence time}\label{sec:samc-convergence}
A primary difficulty in using the SAMC method lies in identifying an
appropriate value for $t_0$.  Although SAMC is proven to formally
converge regardless of the $t_0$ value, a choice that is either too
high or too low will result in prohibitively slow convergence to the true
entropy of the system.  It is instructive to consider separately
values of $t_0$ that are too low or too high.

We can place a rigorous \emph{lower} bound $t_{\min}$ on the number of
moves required to find the true entropy by considering the total change
that needs to be made to the entropy:
\begin{align}
  \Delta S_{\text{tot}} \equiv \sum_E S(E) - S_{\min}.
\end{align}
This connects with $\gamma_t$ because the $\ln w(E)$ (our approximation for entropy)
has $\gamma_t$ added to it on each move.  Thus a minimum number of moves
that could possibly result in the true entropy $S(E)$ starting with a flat set of
weights is determined by the total entropy change required.  We can estimate
this number of moves required by summing the $\gamma_t$,
which we can approximate using an integral:
\begin{align}
   \Delta S_{\text{tot}} &= \sum_{t=0}^{t_{\min}} \gamma_t^{\text{SA}} \\
  &\approx t_0 + \int_{t_0}^{t_{\min}} \frac{t_0}{t}dt
  \\
  &= t_0\left(1 + \ln\left(\frac{t_{\min}}{t_0}\right)\right)
\end{align}
Solving for ${t_{\min}}$ we find that
\begin{align}
  {t_{\min}} &= t_0 e^{\frac{\Delta S_{\text{tot}}}{t_0} - 1}
\end{align}
which means that the minimum time to converge grows exponentially
as $t_0$ is made smaller.  You \emph{seriously} don't want to underestimate
$t_0$!

One might reasonably choose to err by selecting a large $t_0$. The rate of
convergence is harder to estimate when $t_0$ is large, but in general
$\gamma_t^{\text{SA}}$ itself forms a lower bound on the accuracy with which the
entropy may be known, with an unknown prefactor which is related to the
coherence time ($t = t_0$) of the Monte Carlo simulation.  Since
$\gamma_t^{\text{SA}}$ is given by $t_0/t$, the time to converge to a given
accuracy is increased in proportion to the ratio by which we overestimate $t_0$.
Thus, while it is exponentially painful to underestimate $t_0$, overestimating
by several orders of magnitude is also not acceptable.  We should note that
these extreme limiting cases do not preclude the possibility that there is a
wide range of $t_0$ values that lead to an acceptable convergence rate.

\section{SAD Algorithm}\label{sec:sad}
The Stochastic Approximation with Dynamic update factor (SAD) method
is a variant of the SAMC Algorithm that attempts to dynamically choose
the modification factor rather than relying on system dependent
parameters such as $t_0$ or $\gamma_0$.  There is an immediate
advantage of such an algorithm where parameters are chosen independent
of system size or type. Each flat-histogram method has unique
advantages and disadvantages.  Wang-Landau and $1/t$-WL require an
energy range for initialization.  SAMC removes this energy range
requirement but requires simulating every possible energy. Our
proposed method SAD requires the user to input $T_\text{min}$, the
lowest temperature of interest, which is an immediate disadvantage of
the method. However, identifying a minimum temperature of interest
$T_\text{min}$ may be easier for a user than determining in advance
an energy range of interest or the unphysical parameter $t_0$.

We set the maximum temperature of interest as $T=\infty$ for a couple
of reasons.  First, when this temperature is reached, the system is
maximally randomized, so allowing infinite temperature further ensures
that the system can surmount an energy barrier of any size.  Second,
the high temperatures tend to be ``easy'' to simulate, and tend to
converge very quickly.  We note, however, that this is under the
assumption that infinite temperature corresponds to a finite energy,
as is the case for configurational sampling.  Were we to include
kinetic energy in a simulation (as in molecular dynamics), or to
examine a system such as a high density Lennard-Jones fluid with an
entropy maximum at an energy that is far from the energy range of
physical interest, then we would need to apply either a maximum
temperature or a maximum energy, either of which could slow down the
rate at which state space is explored.

\begin{figure}
  \includegraphics[width=\columnwidth]{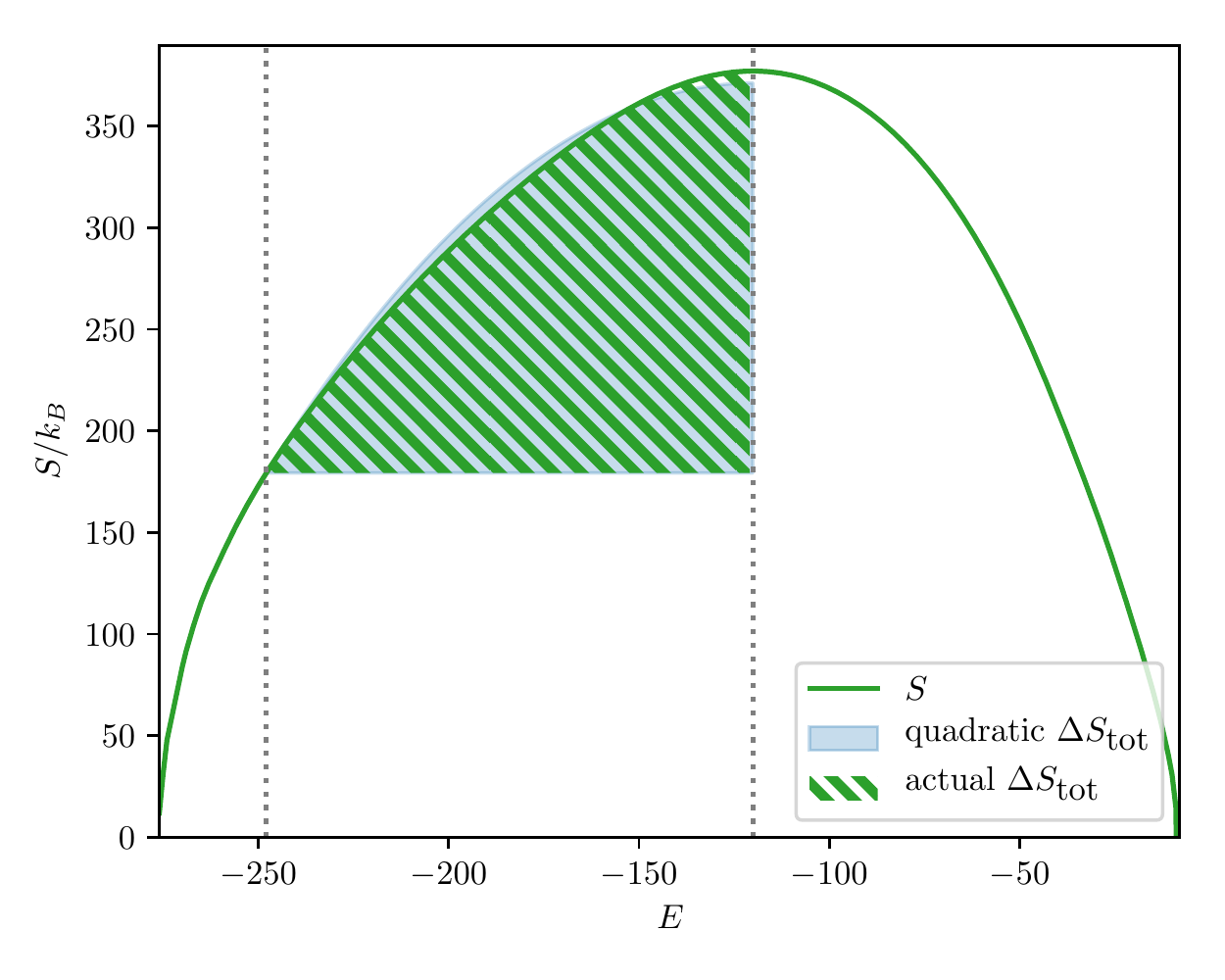}
  \caption{The entropy of a square-well fluid with 50 atoms and filling fraction
        0.3 as the green line.  The green hatched area reflects the
        minimum entropy change needed to converge to the true value.
        The light blue area is the quadratic approximation
        for the change in entropy.  The vertical dotted lines represent
        the energy corresponding to $T=1/3$ and $T=\infty$.}
  \label{fig:entropy-cartoon}
\end{figure}

While for SAMC, the update factor is defined in the original
implementation (see equation~\eqref{eq:1}), for SAD the update factor $\gamma_{t}^{\text{SAD}}$ is
thought of as $\nicefrac{\text{d}S}{\text{d}t}$. This tells us that
the SAMC parameter
$t_0$ should have dimensions of entropy.
We begin with an estimate of the average value of the entropy (relative
to the lowest entropy at $T_{\min}$).  If we assume a quadratic
dependence on energy (see Fig.~\ref{fig:entropy-cartoon}), this is given by
\begin{align}
\langle S\rangle \approx \frac13 \frac{E({T=\infty}) - E(T_{\min})}{T_{\min}}
\end{align}
We approximate this energy difference by $E_H -E_L$ where $E_H$ and
$E_L$ are defined below. The entropy numerator of the update factor in
general should scale with the total number of interesting energy states
$N_S$, since updates to the weights are distributed between that many
energy states.  The product $N_S\langle S\rangle$ is the total change
of entropy required (starting from constant weights) to find the true
entropy, and puts a lower bound on the convergence time. After
\emph{long} times, when all the energies have been explored a long time ago, we
wish for a lower update factor in order to more rapidly refine the
remaining error in entropy.  We track the time at which we first visited
each possible energy.  We define $t_L$ to be the last time that we
encountered an energy that we currently believe is in the energy range
of interest, so a $t\gg t_L$ we feel confident that we have established
the true energy range of interest. We gradually transition to a lower
update factor (but still asymptotically scaling as $\gamma_t \propto
1/t$ to ensure eventual convergence).  Finally, we wish for an update
factor that is \emph{never} greater than 1, because a very large update
factor could introduce very large errors in entropy that may take many iterations
to remove.  The SAD expression for $\gamma_t$ which incorporates these
ideas is:
\begin{align}
  \gamma_{t}^{\text{SAD}} =
     \frac{
       \frac{E_{H}-E_{L}}{T_{\text{min}}} + \frac{t}{t_L}
     }{
       \frac{E_{H}-E_{L}}{T_{\text{min}}} + \frac{t}{N_S}\frac{t}{t_L}
     }
\end{align}
where $E_H$ and $E_L$ are the current estimates for the highest and
lowest energies of interest as defined below.  This factor
asymptotically has the same $1/t$ behavior as the original SAMC
algorithm and with the same $N_S$ prefactor used by the $1/t$-WL
method; however for earlier values of $t$, the update factor drops as
$1/t^2$ and jumps every time a new energy is determined to be of
interest.  This behavior allows SAD to dynamically prevent the update
factor from decreasing too rapidly.

\begin{figure}
  \includegraphics[width=\columnwidth]{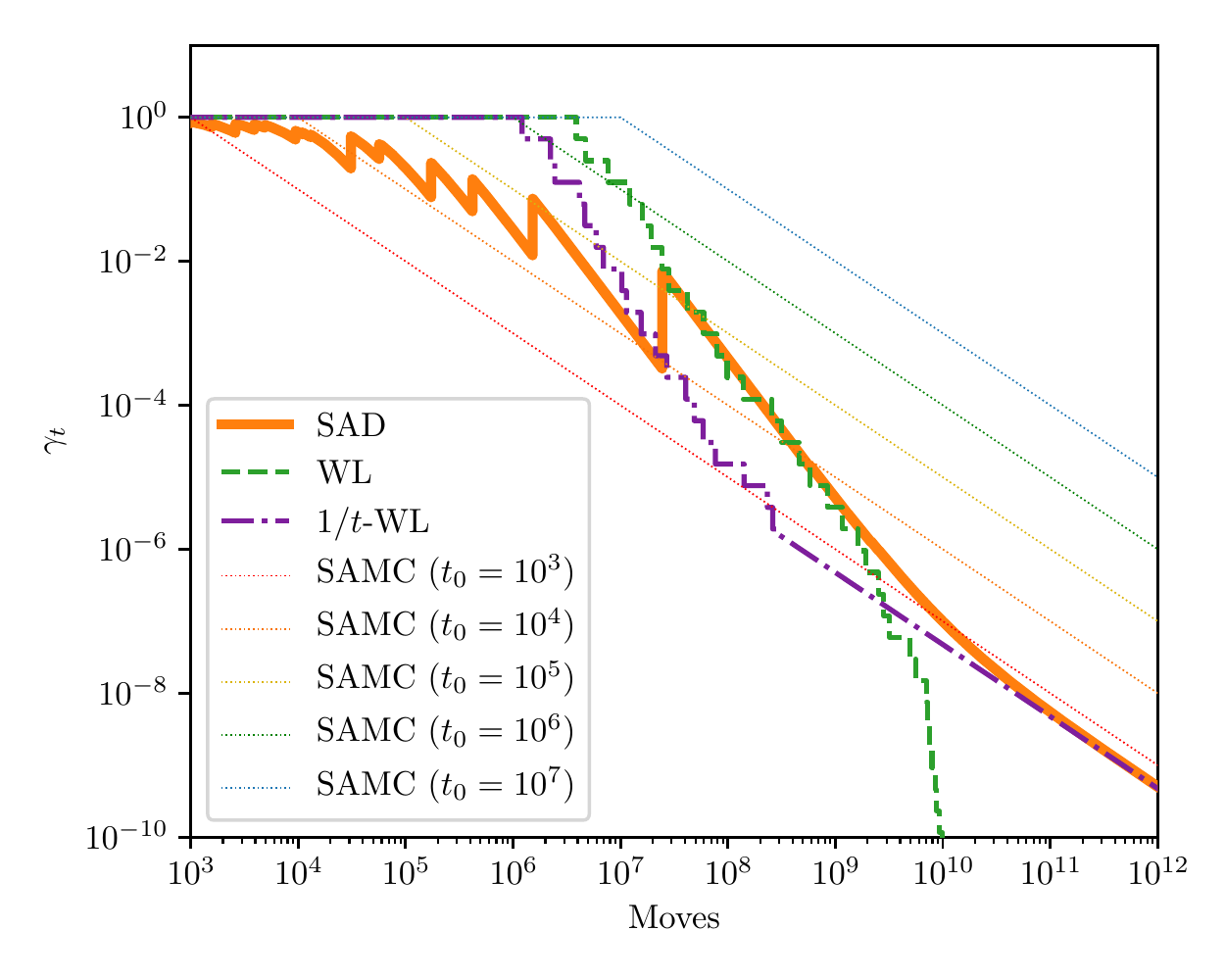}
  \caption{The update factor $\gamma_t$ versus iteration for the
    square-well fluid with 256 atoms, for four different methods: WL, $1/t$-WL, SAMC,
    and SAD.}
    \label{fig:gamma-vs-t}
\end{figure}
Figure~\ref{fig:gamma-vs-t} compares $\gamma_t$ for the related
methods SAD, WL, $1/t$-WL, and SAMC.  For SAMC, $\gamma_t$ remains
constant before dropping as $1/t$.  WL $\gamma_t$ remains at 1 for
many iterations, and then decreases very rapidly, with $1/t$-WL
behaving similarly but decreasing more aggressively before
transitioning to a more conservative $1/t$ behavior.  The update
factor for SAD fluctuates dynamically around a value less than 1 for
early MC moves, and then decreases as approximately $1/t$ while
continuing to fluctuate as new energies are found to be important.  At
intermediate times, the SAD $\gamma_t$ decreases as $1/t^2$ before
asymptoting to $N_S/t$, which is the same as $1/t$-WL.

Since SAD does not explore \emph{all} energy states, it needs to
determine what energy range corresponds to the temperature range of
interest defined by $T_{\min}<T<\infty$. The simulation is responsible
for determining and updating this energy range. Given the true entropy
$S(E)$, we can define the interesting energy range as
  $E(T_{\min}) <E< E(T=\infty)$
where $E(T)$ is the energy that maximizes $S-E/T$.  During the course of the
simulation, this precise energy is challenging to evaluate accurately.
In order to ensure that we sample this entire energy range adequately,
we define two energy limits:  a high energy $E_H$ and a low
energy $E_L$, which define the range over which the energy histogram
is made flat. At move $t$, $E_H$ and $E_L$ are the greatest and lowest
energy that prior to that move
had the highest histogram value (i.e. been visited the most times) at some point
during the course of the simulation.
This definition results in a ``ratcheting'' effect, in which $E_H$
may only increase, while $E_L$ may only decrease over the course of the
simulation, which results in a conservative estimate of the range of
energies that need be sampled.

During the simulation when considering a move inside the energy
range of interest $E_L \le E \le E_H$, the weights are used as in the
three methods already described.  If $E\ge E_H$, the weight is taken
to be
\begin{align}\label{eq:highw}
  w(E>E_H) &= w(E_H),
\end{align}
which corresponds to an infinite temperature.  This choice ensures
that if the maximum in entropy is at an energy $E_{\max}>E_H$, then
the energy $E_{\max}$ will eventually have the highest number of
counts and the ratcheting will result in $E_H\ge E_{\max}$ .  At lower
energies, Boltzmann weights corresponding to the minimum temperature
are used:
\begin{align}\label{eq:loww}
  w(E<E_L) &= w(E_L)e^{-\frac{E_L-E}{T_{\min}}}.
\end{align}
This choice has the result that if the energy $E_{\min}$ at which the free
energy at $T_{\min}$ is minimized is less than $E_L$, the lower energy
limit will ratchet down to include $E_{\min}$.
Each time we change the value of $E_H$ or $E_L$, the weights outside the
new portion of the interesting energy range are set to the expressions
in Equations~\ref{eq:highw} and~\ref{eq:loww}.


A significant advantage of SAD over SAMC---which the $1/t$-WL and WL
methods share \emph{after} they have discovered all the energies---is that
the schedule for $\gamma_t$ automatically responds to the choice of bin
size.
SAD should perform similarly over a reasonable range of bin sizes
because $\gamma_t \propto \nicefrac{N_S}{t}$.  As the number of energy
states $N_S$ found increases (fine binning), the time spent $t$ in
each bin will decrease with the effect that the convergence should be
roughly independent of the bin size chosen.  SAMC could be used with a
prefactor $\gamma_0$ to aid in a similar
way~\cite{werlich2015stochastic} but this adds yet another parameter
for the user to choose.

\section{Results}\label{sec:results}

\subsection{Square-well fluid}
As our first test case, we consider the square-well fluid i.e. a
system of particles whose interactions are governed by a square-well
potential~\cite{singh2003surface, barker2004perturbationSW}.  The
square-well potential is an ideal test-bed as it is a simple model for
a liquid, which includes both attractive and repulsive
interactions~\cite{barker1967-SW-perturbation, vega1992phase}.
To date, there have not been any published direct convergence comparison tests
for flat histogram methods applied to the square-well fluid.
The potential $U(\textbf{r})$ for such a system is given by
\begin{equation}
 U(\textbf{r})=\begin{cases} \infty &
 \lvert\textbf{r}\rvert< \sigma\\-\epsilon &
 \sigma<\lvert\textbf{r}\rvert<\lambda\sigma\\0 &
 \lvert\textbf{r}\rvert > \lambda\sigma\end{cases}
\end{equation}
where $\sigma$ is the hard-sphere diameter of the particle, $\lambda$ is the
reduced range of the potential well, and $\epsilon$ is its depth. This model has
the further advantage that binning is not required because the energy is
discrete.

\begin{figure*}
  \includegraphics[width=0.5\textwidth]{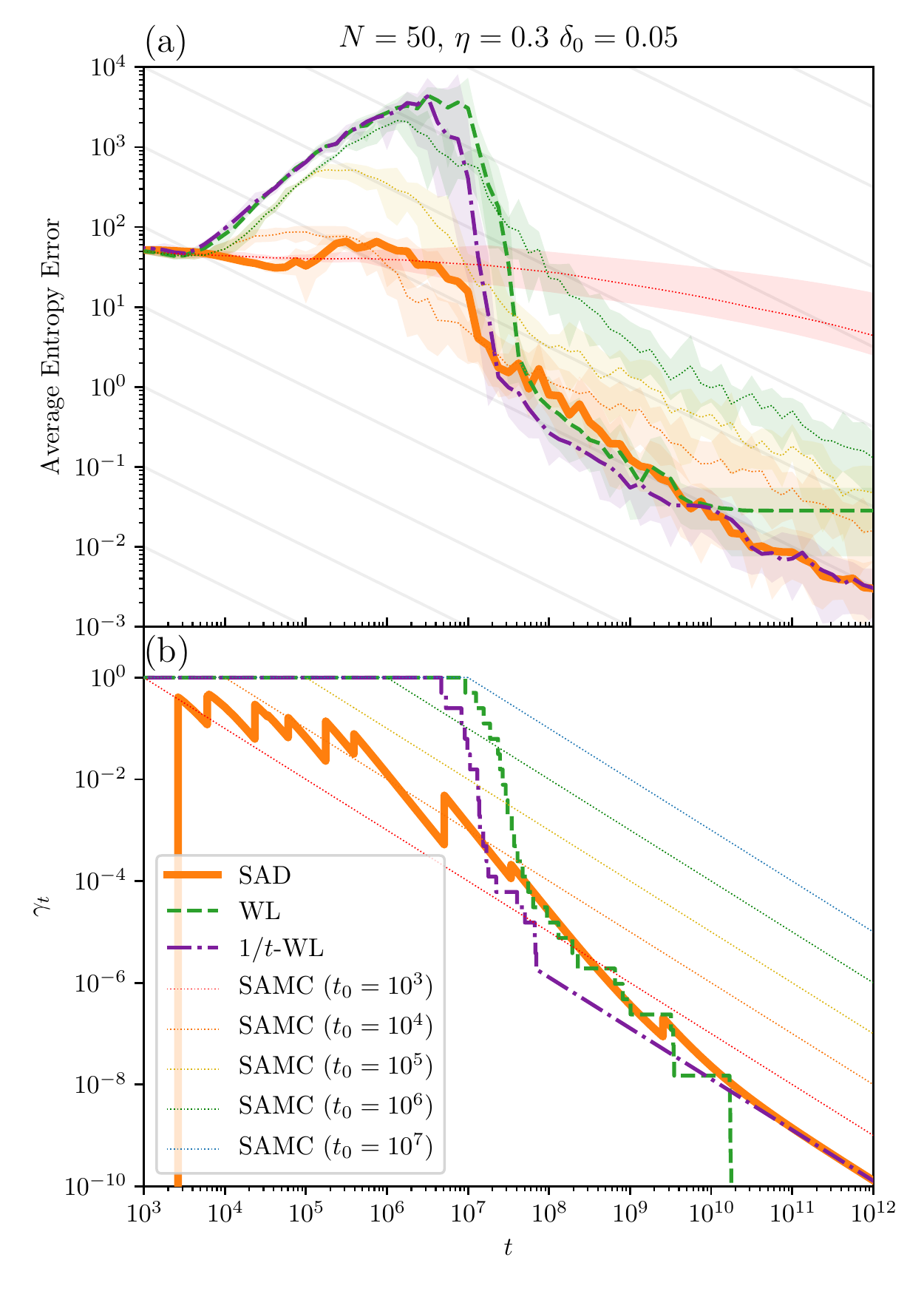}%
\includegraphics[width=0.5\textwidth]{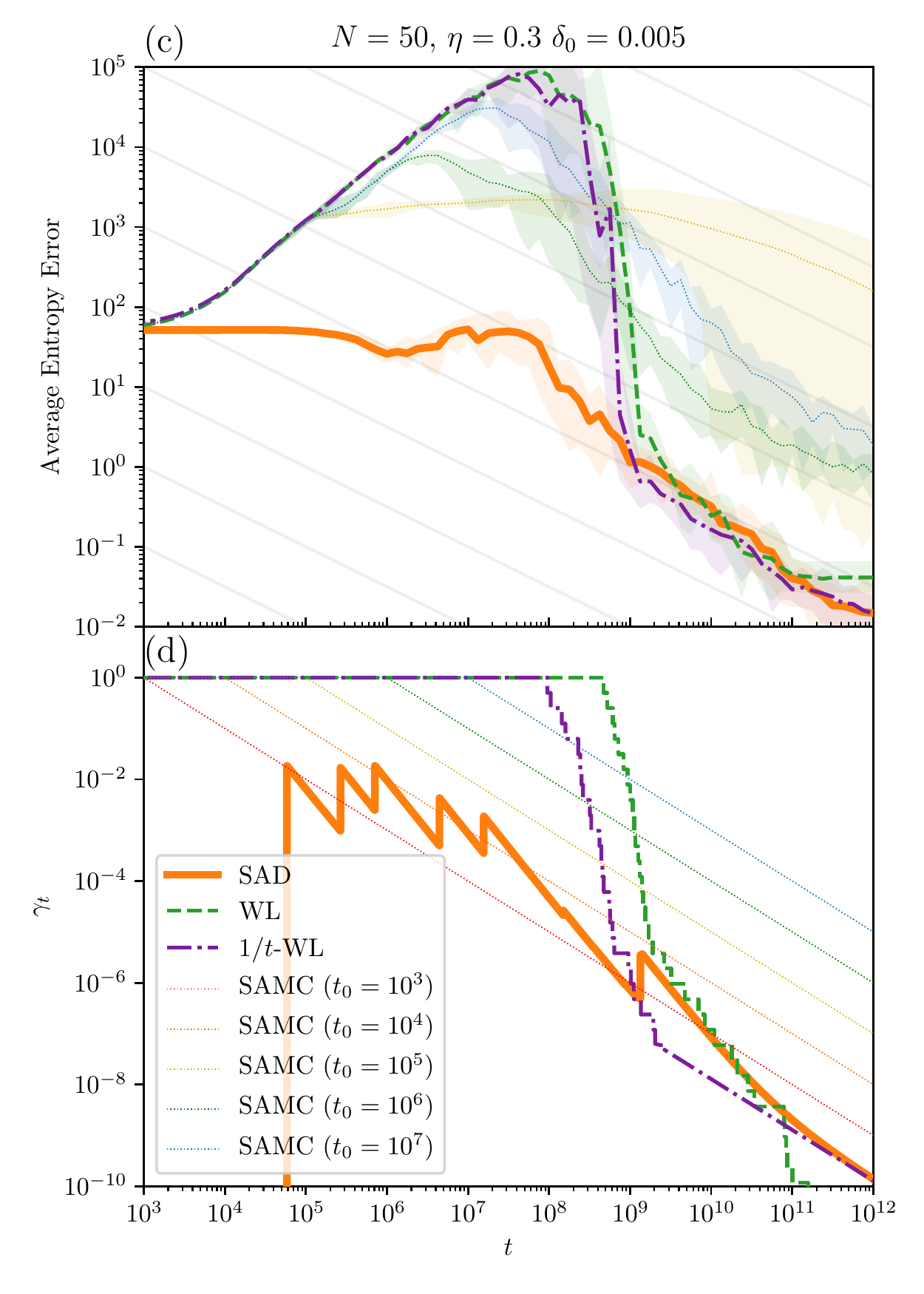}
  \caption{(a) The average entropy error for each MC method for $N=50$,
               $\delta_0 = 0.05\sigma$, $\eta = 0.3$, and $T_{\min} = 1/3$
               as a function of number of iterations run. The error is
               averaged over 8 independent simulations, and the best
               and worst simulations for each method are shown as a
               semi-transparent shaded area, and
           (b) the update factor $\gamma_t$ versus iteration number
               for the same simulations.
           (c) The average entropy error for each MC method for the
               same physical system with a smaller displacement distance
               $\delta_0 = 0.005\sigma$,
               as a function of number of iterations run, and
           (d) the update factor $\gamma_t$ versus iteration number
               for the same simulations.
  }\label{fig:fast-slow-gamma}
\end{figure*}

We tested the algorithms on two square-well fluid systems.  The first
is a smaller simulation with a particle number of 50, a well-width of
$\lambda = 1.3$, and a volume corresponding to a filling fraction
(defined as the fraction of volume filled by atoms) of $\eta =
0.3$. The second system is larger, with a particle number of 256, a
well-width of $\lambda = 1.5$, and a volume corresponding to a filling
fraction of $\eta = 0.17$.  For each system we use a reasonable
root-mean-square displacement distance $\delta_0 = 0.05\sigma$ for
proposed moves, and for the smaller system we also use an unreasonably
small displacement distance of $0.005\sigma$. The simulations explore
the energy space of the systems with minimum reduced temperatures of
$T_{\text{min}} = 1/3$ for simulations of the smaller system, and
$T_{\min}=1$ for the larger system.  All simulations lead to the
minimum important energy $E_{\min}$ and maximum entropy energy
$E_{\max}$ being calculated (with the exception of the WL methods
where both of these parameters are needed \emph{a priori}).

The SAMC simulations
computed the density of states for the entire range of possible
energies.  The SAD simulations determined the energy range of interest
dynamically as described above, based on a specified $T_{\min}$.  For
the WL and $1/t$-WL simulations, we constrained the simulation to
remain in the energy range corresponding to $T_{\min} < T < \infty$,
as determined by a previous SAMC simulation.  Thus the WL and $1/t$-WL
simulations were given extra information that in practice would not be
available without additional computational effort, and the SAMC simulations
computed the entropy over the entire range of possible energies, which
required more effort.

We use the average entropy error versus moves as a metric to compare
simulation runtimes and overall convergence. The overall accuracy
is determined by examining the fractional error of a particular method to
a precise reference system. For each simulation, the reference system
is chosen to be the final output of a SAMC simulation with a fixed energy range
corresponding to the temperature range of interest.
Although SAMC does not require an energy range as an input parameter,
we find that by limiting the simulation to this energy range, we can
achieve much faster convergence with a smaller $t_0$.  We compute an average
of the error by averaging the error in the entropy over the
interesting energy range, and then averaging \emph{this} error over several
simulations run with different random number seeds.

\subsubsection{50 square-well atoms}
For this 50 atom simulation, we chose a minimum reduced
temperature of $1/3$, which corresponds to an interesting energy range
from $-248$ to $-120$.  The number of important energy states
for this system is therefore $N_S = 129$.  The entropy of this system is shown in
Fig.~\ref{fig:entropy-cartoon} above, which shows that over this
energy range the entropy differs by 198, corresponding to a ratio of
$10^{86}$ between the highest and lowest density of states.

In order to explore the effect of simulation details on convergence,
we consider two values for the displacement distance by which atoms
are moved during a Monte Carlo step.  We began with a reasonable
displacement distance of $\delta_0 = 0.05\sigma$, which corresponds to
an acceptance rate of proposed moves of 38\%.  We further ran
simulations with a much smaller displacement distance of $\delta_0 =
0.005\sigma$, which resulted in an acceptance rate of 86\%, which
converged more slowly.

Figure~\ref{fig:fast-slow-gamma}a shows the average error in the
entropy as a function of time for this system with the reasonable
displacement distance of $\delta_0 = 0.05\sigma$.  The solid/dashed lines
represent the average of the absolute value of the error in the
entropy averaged over eight simulations using different random number
seeds.  The range of average errors for each simulation is shown as a
shaded region around its mean error.  By the time $10^8$ moves have
been made all but the SAMC simulation with the shortest $t_0$ have
begun to converge as $1/\sqrt{t}$.  We then see the WL error
saturate around $10^{10}$ moves.

Figure~\ref{fig:fast-slow-gamma}c shows the average error in
the entropy as a function of time for this system with the
unreasonably small displacement distance of $\delta_0 = 0.005\sigma$.
The smaller translation scale causes all methods to take additional time to
explore all energies. Based on random walk scaling, the convergence time of an
ideal method should scale roughly as $\delta_0^{-2}$ in the limit of small
$\delta_0$, that is, one order of magnitude decrease in the displacement
distance should result in two order of magnitude increase in convergence time.
SAMC simulations with a $t_0$ value that rapidly converged for $\delta =
0.05\sigma$ do not converge at all in $10^{12}$ moves for a translation scale of
$\delta = 0.005\sigma$. It is also worth noting that for the smaller
displacement distance, the SAMC rule of thumb of choosing $t_0$ to be
approximately $100N_S$ is no longer valid. SAD, WL, and $1/t$-WL handle the
shift in displacement distance and converge roughly as expected.

The methods SAD, WL, and $1/t$-WL compensate for the smaller
displacement distance by reducing $\gamma_t$ more slowly, as can be seen from
Figure~\ref{fig:fast-slow-gamma}b and ~\ref{fig:fast-slow-gamma}d.
The update factors take approximately 10$\times$ longer to reach steady-state
for the smaller displacement distance. Because of this update behavior,
these methods are less sensitive to the choice of displacement
distance than SAMC is.

%

\begin{figure}
\includegraphics[width=\columnwidth]{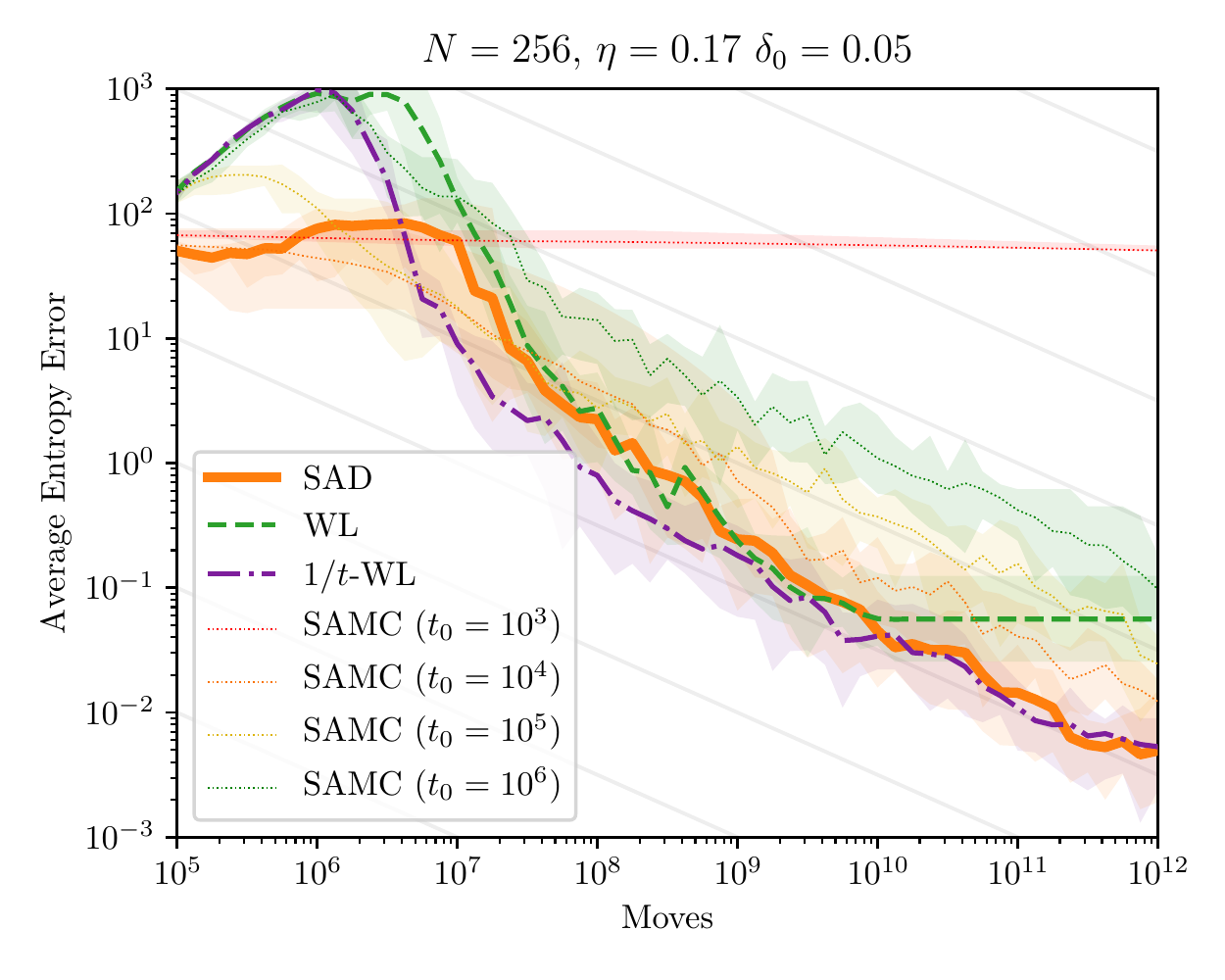}
  \caption{
  The average entropy error for each MC method for $N=256$,
               $\delta_0 = 0.05\sigma$, $\eta = 0.17$, and $T_{\min} = 1$
               as a function of number of iterations run.  The error is
               averaged over 8 independent simulations, and the best
               and worst simulations for each method are shown as a
               semi-transparent shaded area.  The update factor for this
               system is in Fig.~\ref{fig:gamma-vs-t} above.}\label{fig:n256}
\end{figure}

\subsubsection{256 square-well atoms}
Next we introduce a considerably larger simulation containing 256
atoms which has a maximum entropy about 1500 greater than its minimum.
This makes exploring the entire range of energies extremely expensive,
and strongly favors the methods that restrict the energy (or
temperature) range of interest.  For this simulation, we chose a much
higher minimum reduced temperature of $1.0$, which corresponds to an
interesting energy range from $-915$ to $-509$.  The number of
important energy states for this system is then $N_S = 407$.  The
minimum entropy over this energy range is just 395 less than the
maximum, corresponding to a ratio of only $10^{118}$ between the
highest and lowest density of states.


Figure~\ref{fig:n256} shows the average error in the entropy as a
function of moves for this system with the reasonable displacement
distance of $\delta_0 = 0.05\sigma$.  The solid lines represent the
average of the absolute value of the error in the entropy averaged
over eight simulations using different random number seeds.  The range
of average errors for each simulation is shown as a shaded region
around its mean error.  By the time $10^8$ moves have been made all
but the SAMC simulation with the shortest $t_0$ shown have begun to
converge as $1/\sqrt{t}$.  We then see the WL error saturates around
$10^{10}$ moves.  Once again, the convergence of SAD is essentially
the same as that of the $1/t$-WL method.

\begin{figure}
  \includegraphics[width=\columnwidth]{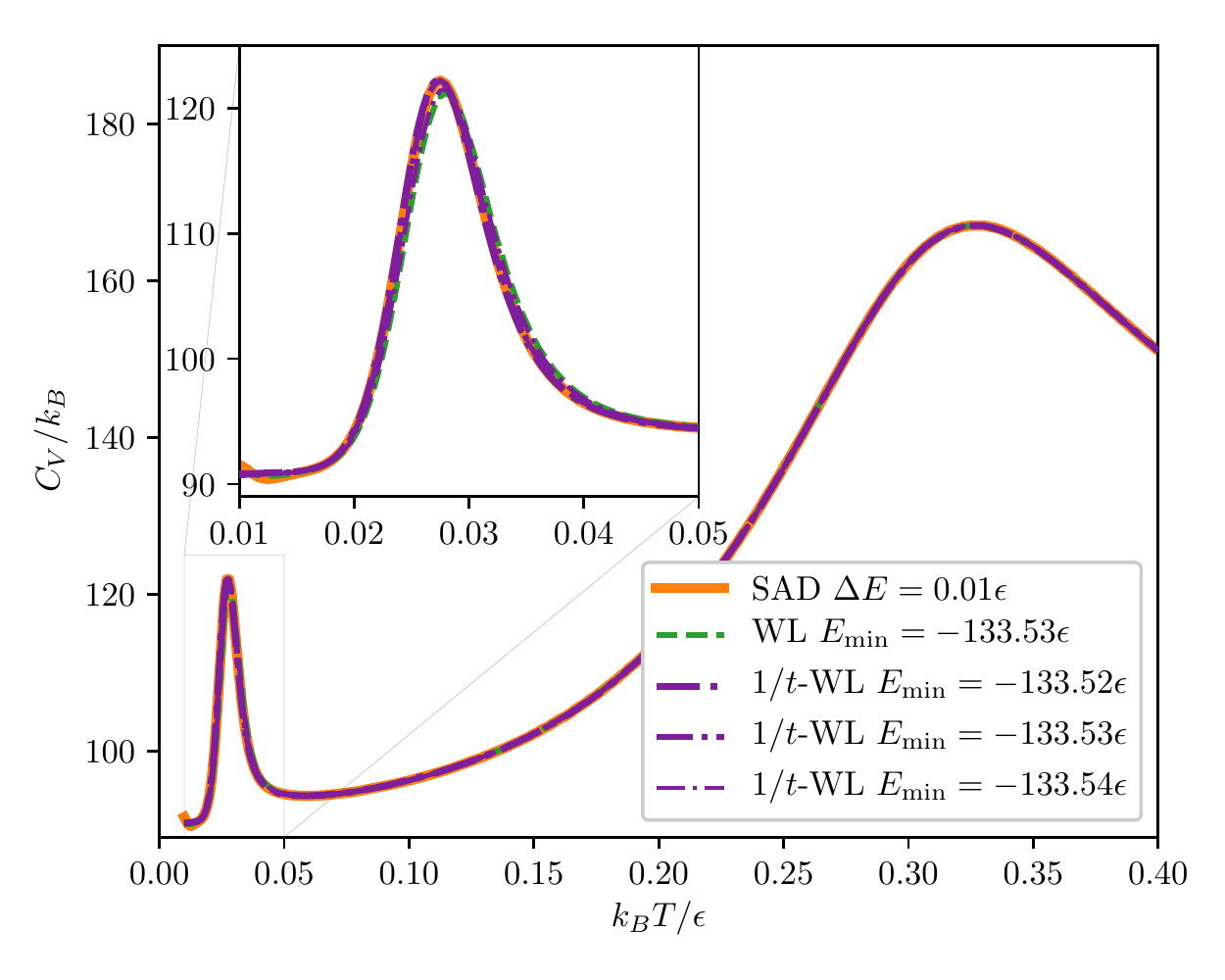}
  \caption{The heat capacity of the LJ31 cluster after $10^{12}$ moves
    as calculated with each method.  Each simulation used a bin size
    of $\Delta E=0.01\epsilon$.}
  \label{fig:lj-cv}
\end{figure}

\subsection{31-atom Lennard-Jones cluster}

We also examined convergence of each algorithm on the 31-atom cluster
(LJ31), which is the smallest Lennard-Jones cluster that exhibits
broken ergodicity and a solid-solid phase transition at a low
temperature~\cite{doye1998thermodynamics,
  martiniani2014superposition}.  Clusters of Lennard-Jones atoms have
been frequently used for testing Monte Carlo
algorithms~\cite{wales1997global, neirotti2000phase,
  frantsuzov2005size, mandelshtam2006multiple}.  The LJ31 cluster
is commonly used for testing convergence of algorithms
because it features a two-funnel energy landscape with a significant
barrier between the two low-energy states~\cite{calvo2000entropic,
  calvo2000phase, poulain2006performances,
  martiniani2014superposition}.  In particular, the LJ31 cluster
features a low-temperature solid-solid phase transition that is challenging to
converge~\cite{mandelshtam2006multiple}.  Poulain \emph{et al.}
tested a number of variants on the WL algorithm on LJ31 and failed to
converge the spike in heat capacity corresponding to the solid-solid
phase transition~\cite{poulain2006performances}.


Figure~\ref{fig:lj-cv} shows the heat capacity of the LJ31 cluster
over the temperature range from $k_BT=0.01\epsilon$ to
$k_BT=0.40\epsilon$.  There are two peaks, a large peak around $k_BT
\approx 0.3\epsilon$ corresponding to melting temperature, and a
small peak around $k_BT \approx 0.027\epsilon$ for the solid-solid
transition from a Mackay-to-anti-Mackay (M $\rightarrow$ aM)
transition~\cite{doye1997thermally, hendy2001molecular,
  mandelshtam2006multiple}.  We constrain our atoms within a spherical
box of radius 2.5$\sigma$, as is common in the
literature~\cite{poulain2006performances, mandelshtam2006multiple}.
We note that using a larger box (with radius $5\sigma$) has a large
effect on the melting peak, and a smaller but still significant effect
on the solid-solid transition.  Since our focus is on the convergence
rate, we will restrict ourselves to the smaller box size.  In addition, we
further restricted all our simulations to negative total potential energies.

\begin{figure}
  \includegraphics[width=\columnwidth]{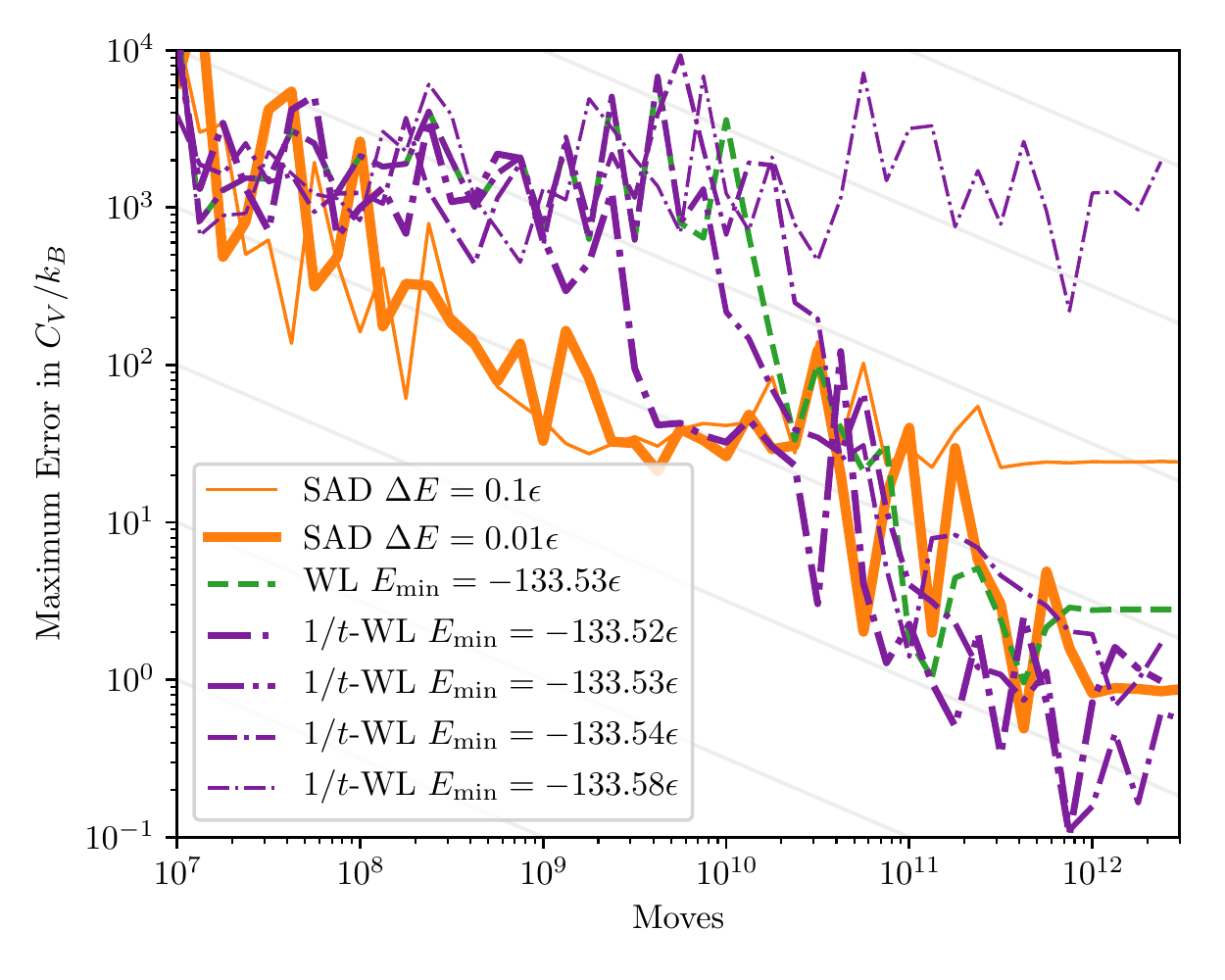}
  \caption{The maximum error in the heat capacity of the LJ31 cluster
    in the temperature range 0.01~$\epsilon\le k_BT \le 0.05\epsilon$.
    Note that the value of $C_V$ ranges from 90$k_B$ to 120$k_B$, so
    the smallest (maximum) errors seen are around 1\%.  All WL methods
    shown were run with a histogram bin width of $\Delta
    E=0.01\epsilon$.  The converged reference heat capacity was
    computed with a $1/t$-WL simulation with the energy constrained to
    a range $-133.53\epsilon\le E\le -110\epsilon$.
  }
  \label{fig:lj-cv-error}
\end{figure}

In order to test the low-temperature convergence, we focused on the
heat capacity in the temperature range $0.01\epsilon \le k_BT \le
0.05\epsilon$ (see inset in Fig.~\ref{fig:lj-cv}), which contains
the solid-solid phase transition.
We compute the heat capacity from the entropy using a standard
canonical analysis, which gives
\begin{align}
  C_V &= 31\times\frac32k_B + \frac1{k_BT^2}\left< (E- \langle
  E\rangle)^2\right>
  \label{eq:cv}
\end{align}
where the first term represents the kinetic energy contribution, and
the second term is the potential energy contribution, with canonical
averages using the microcanonical entropy to compute multiplicities.
For our reference heat capacity, we ran a $1/t$-WL simulation with the
energy constrained to a range $-133.53\epsilon\le E\le -110\epsilon$.
We chose this energy range through experimentation, as we found the
convergence rate to be very sensitive to the minimum energy chosen.

We ran simulations with several bin widths using each algorithm we studied
and show results for $\Delta E=0.1\epsilon$ and
$0.01\epsilon$.
When examining the converged results, we saw that each bin size yielded a
measurably different low temperature heat capacity. The largest
bin size of $0.1\epsilon$ introduced a maximum error in heat capacity
of more than $20k_B$ compared with the smaller bin size.  We will focus on
the smaller bin size, and
used a bin size of $0.01\epsilon$ for our converged reference heat capacity.

Figure~\ref{fig:lj-cv-error} shows the maximum error in heat capacity
over the temperature range $0.01\epsilon\le k_BT\le 0.05\epsilon$ as a
function of the number of Monte Carlo moves performed.
We explored the convergence behavior of each algorithm at two energy
bin widths ($\Delta E=0.01\epsilon$ and $\Delta E=0.1\epsilon$).  We
were unable to converge the WL or $1/t$-WL algorithms with the larger
bin size.
The rate of convergence of SAD for the heat capacity seems to be roughly
independent of the bin width.  In contrast (but not shown) the entropy
itself shows much larger statistical fluctuations with smaller bin
sizes.  The canonical averaging process in Eq.~\ref{eq:cv} averages
out those statistical fluctuations, which reduces the penalty of using
a smaller bin size.


The convergence behavior of the WL and $1/t$-WL is incredibly
sensitive to the minimum energy chosen, which complicates the process of
obtaining a workable result.  The known ground state energy of this
cluster is $-133.586422\epsilon$~\cite{northby1987structure,
  wales1997global}.  With a minimum energy of $-133.58\epsilon$, none
of our WL or $1/t$-WL simulations converged in 10$^{12}$ moves.  With minimum
energies of $-133.54\epsilon$, $-133.53\epsilon$, and
$-133.52\epsilon$, the simulations do seem to converge, although the
lowest energy case was significantly slower.
This result highlights the advantage of being able to
specify a temperature range rather than an energy range when computing
properties as a function of temperature.  It is not only more
convenient, but also far more efficient due the the fact that only a
single simulation need be performed.


Finally, we want to mention our results for SAMC, which are not shown
in our plots in order to reduce clutter.  With a well-chosen power of ten value for
$t_0$, SAMC can converge to the correct result, but less rapidly than
either $1/t$-WL or SAD.  However, with a poor $t_0$
parameter, the method cannot converge in practice.  In addition, the
optimal $t_0$ parameter depends on the bin width as well as the energy
range chosen, further complicating matters.


\section{Conclusions}
We have introduced a new algorithm, a variant of the SAMC method, which
effectively samples the
energy space corresponding to a desired range of temperatures for a
few systems.
We find that both SAD and $1/t$-WL converge more rapidly than SAMC, and
unlike WL consistently converge to the correct density of states. SAD requires
the user to specify a temperature range of interest rather than an
energy range of interest as $1/t$-WL does.  For use cases in which
a range of desired temperatures is known, this will make the SAD method
considerably more convenient.

We find that SAMC converges for a reasonable choice of
$t_0$ but this parameter can be difficult to tune especially across
significantly differing systems.  We find that even simple
changes to the Monte Carlo moves can have a dramatic effect on the
range of practical $t_0$ values.  Additionally, SAMC does not
converge as rapidly as either SAD or $1/t$-WL even for the best choice of $t_0$,
when a relatively small range of energies is required, because it always
simulates all possible energies.

Finally, we find that for examining a Lennard-Jones cluster the WL and
$1/t$-WL algorithms---versions of which had previously been found to
be unsuccessful on this problem~\cite{poulain2006performances}---have
convergence properties that are highly dependent on the choice of
energy range to be examined.  This is particularly problematic when
examining the low-temperature heat capacity, where it is difficult to
determine the lowest energy that will have a significant impact at the
temperatures of interest. This is an example of how SAD is more
powerful when the range of energies is not known in advance.



\bibliography{paper}

\end{document}